\documentclass[preprint,aps,superscriptaddress,nofootinbib]{revtex4}

\usepackage{amsmath}
\usepackage{graphicx}
\usepackage{color}
\usepackage{slashed}

\begin{document}

\title{Offshell thermodynamic metrics of the Schwarzschild black hole}
\author{Wen-Yu Wen}\thanks{%
E-mail: steve.wen@gmail.com}

\affiliation{Department of Physics and Center for High Energy Physics, Chung Yuan Christian University, Chung Li City, Taiwan}
\affiliation{Leung Center for Cosmology and Particle Astrophysics\\
National Taiwan University, Taipei 106, Taiwan}

\begin{abstract}
Thermodynamic metric usually works only for those black holes with more than one conserved charge, therefore the Schwarzschild black hole was excluded.  In this letter, we compute and compare different versions of {\sl offshell} thermodynamic metric for the Schwarzschild-like black hole by introducing a new degree of freedom.   This new degree of freedom could be the running Newton constant,  a cutoff scale for regular black hole, a noncommutative deformation, or the deformed parameter in the nonextensive Tsallis-R{\`e}nyi entropy.  The {\sl onshell} metric of the deformed Schwarzschild solution would correspond to the submanifold by gauge fixing of this additional degree of freedom.  In particular, the thermal Ricci scalar for the Schwarzschild black hole, though different for various deformation, could be obtained by switching off the deformation.
\end{abstract}

\maketitle



\section{Introduction}

The thermodynamic metric has been studied for decades.  First concept was brought by Weinhold \cite{Weinhold:1975}, who regarded the Hessian of internal energy as a mretic of quilibrium states.  Then there was the Ruppeiner metric \cite{Ruppeiner:1979} given by the negative Hessian of entropy with respect to conserved charges.  It was shown that the Weinhold and Ruppeiner are conformally equivalent to each other \cite{Salamon:1984}.  Quevedo suggested a Legendre invaraint form of metric in the formalism of geometrothermodynamics\cite{Quevedo:2006xk, Quevedo:2007ws} and Hendi et al \cite{Hendi:2015rja} proposed another new metric such that divergence of Ricci scalar coincides with phase transition points. 

For the black hole with single charge, such as the Schwarzschild black hole, the charge has to be its ADM mass $M$ and represents the internal energy of the system.  According to the first law of thermodynamics, its mass can be expressed as a function of the Bekenstein-Hawking entropy $S$, say $M=M(S)$, which is a monotonically increasing function of single varaible and $S$ could be seemed as a reparametrization of a one-dimensional curve for $M\ge 0$.  The thermodynamic metric, of course, is trivial for such a one-dimensional space.  However, a theory of quantum gravity is expected to generate a ultraviolate (UV) scale, denoting  $\Lambda_X$, to prevent the catastrophe of infinite temperature predicted by the Schwarzschild metric at its final stage.   Although satisfactory theory of quantum gravity is still unavailable, it may be still possible to construct an effective theory consistent with the General Relativiry (GR) but with the desired UV scale.  This effective theory should correspond to a modified Schwarzschild metric and one may assign a one-parameter deformation function $M=M_X(S)$, where deformation parameter $X$ is determined by the desired UV scale $\Lambda_X$.  If we regard $X$ to be a new degree of freedom, meaning that scale $\Lambda_X$ is allowed to run freely, we obtain an energy function of two variables instead, say $M=M(S,X)$.  In this way, we may have a nontrivial two-dimensional thermodynamic metric and the modified  Schwarzschild-like solution is just the submanifold after a specific $X$ is chosen.  As an analogy to the gauge theory, one may regard this two-dimensional thermodynamic metric offshell and the onshell effective theory is obtained by gauge fixing $X$.  In this letter we will focus on four different effective theories: the Schwarzschild solution with a running Newton's gravitational constant \cite{Falls:2012nd}, the regular Schwarzschild-like black hole with a UV cutoff \cite{Hayward:2005gi}, the noncommutative geometry inspried Schwarzschild black hole \cite{Nicolini:2005vd}, and the Schwarzschild black hole with a nonextensive Tsallis entropy\cite{Biro:2013cra}.

This paper is organized as follows: in the section II, we studied various thermodynamic metric for the Schwarzschild black hole with a running Newton's gravitational constant $G$.  We found it give vanishing Ricci scalar if one simply regards $G$ as a free parameter.  In stead, if one regarded the UV scale on which $G$ depends as the new degree of freedom, then nontrivial Ricci scalar can be obtained.  In the section III, we studied various thermodynamic metric for the regular Schwarzschild-like black hole and regarded the UV length scale as the free parameter.  Zeros of heat capacities and poles in the Ricci scalars in each thermodynamic metric were computed and compared.  In the limit of ordinary Schwarzschild black hole, most Ricci scalars are found inversely proportional to the entropy to some power.  In the section IV, we studied various thermodynamic metric for the noncommutative geometry inspired Schwarzschild black hole.  We computed and compared the zeros and spikes in the heat capacities with the poles and spikes in the Ricci scalars.  The heat capacities bahave like the ordinary Schwarzschild black hole for large entropy, while the Ricci scalar was found inversely proportional to the noncommutative parameter at the extremal limit where the Hawking temperature vanishes.  In the section V, we studied various thermodynamic metric for the Schwarzschild black hole but with the Tsallis-R{\`e}nyi entropy.  While the thermal Ricci scalars in the Ruppeiner and Weinhold behaved  similarly as those in the running Newton constant, they behaved very differently in the HPEM metric. We summarize our result in the section VI.

\section{Shcwarzschild black hole with running Newton constant}
\begin{figure}
\includegraphics[width=0.48\textwidth]{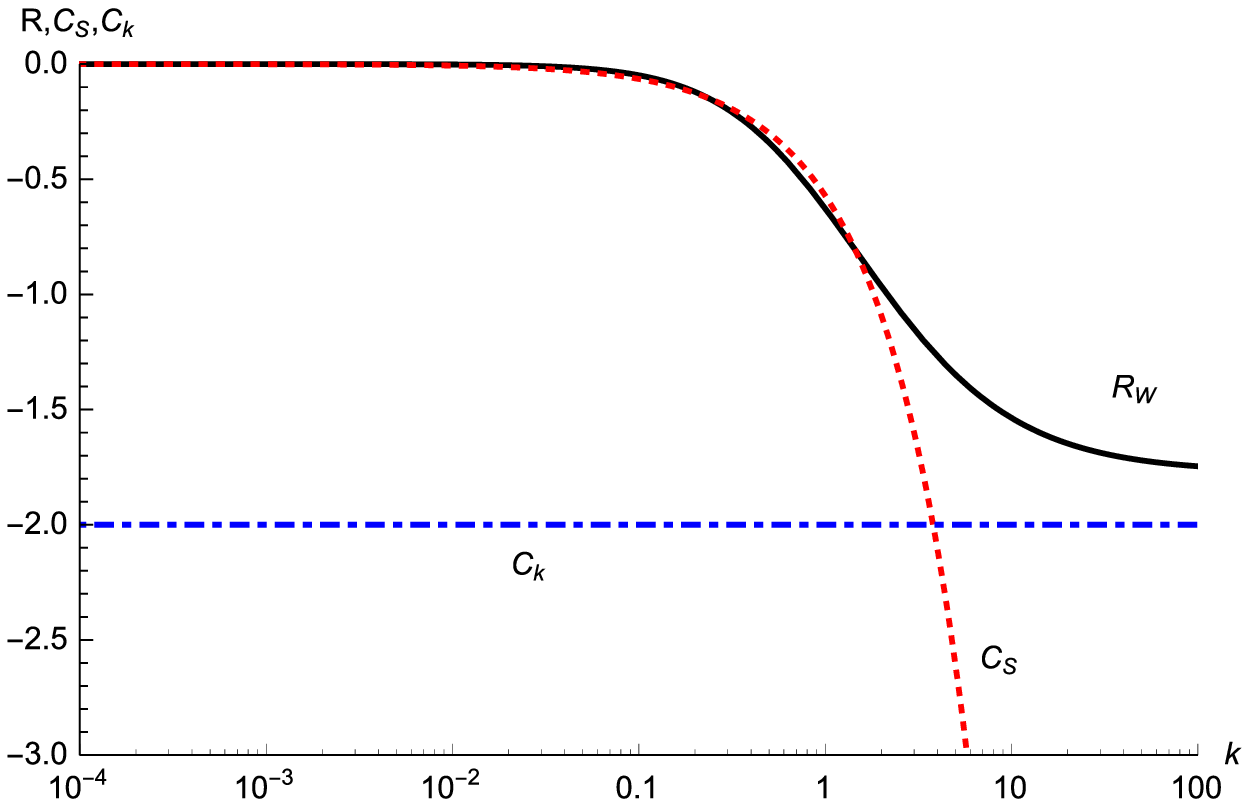}
\includegraphics[width=0.48\textwidth]{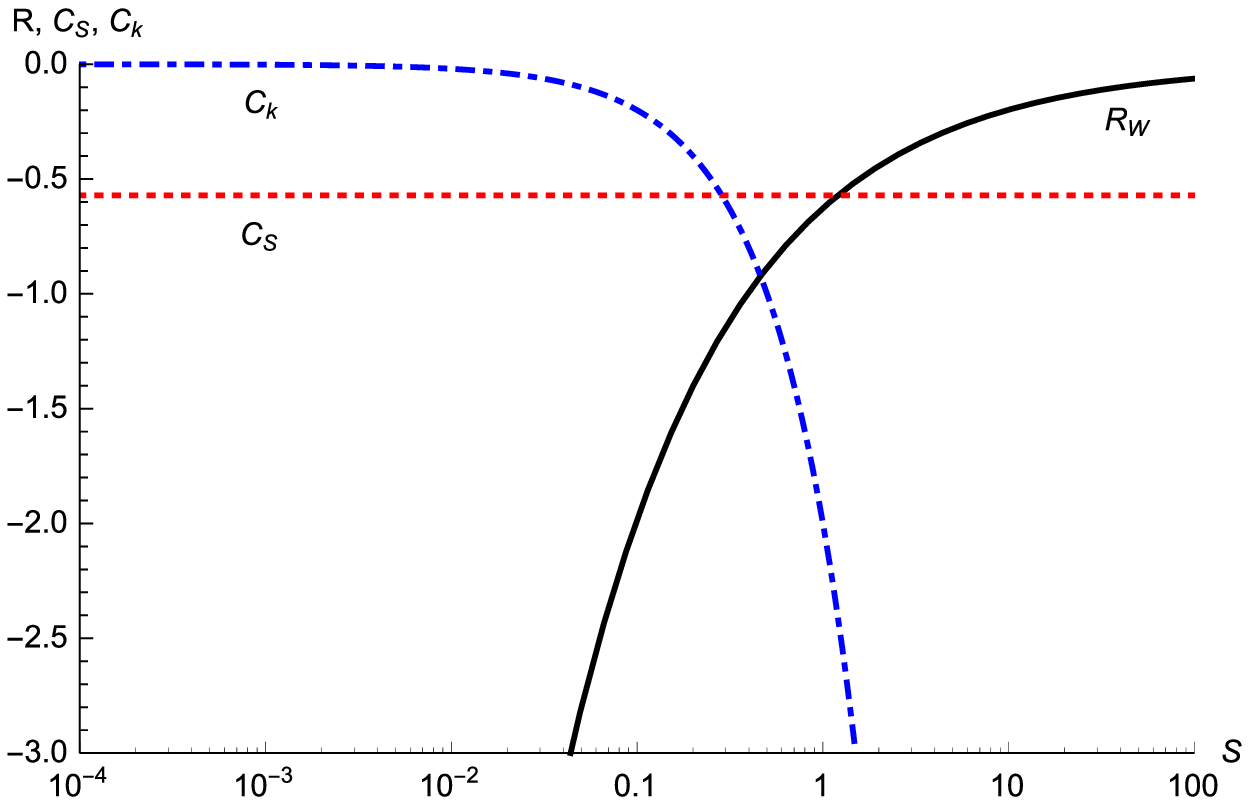}
\caption{\label{fig:Weinhold_runG}(Left) Ricci scalar $R_W$ and heat capacities $C_{k}$ and $C_S$ for various $k$ at fixed $S=1$. (Right) Ricci scalar $R_W$ and heat capacity $C_{k}$ and $C_S$ for various $S$ at fixed $k=1$.}
\end{figure}

\begin{figure}
\includegraphics[width=0.48\textwidth]{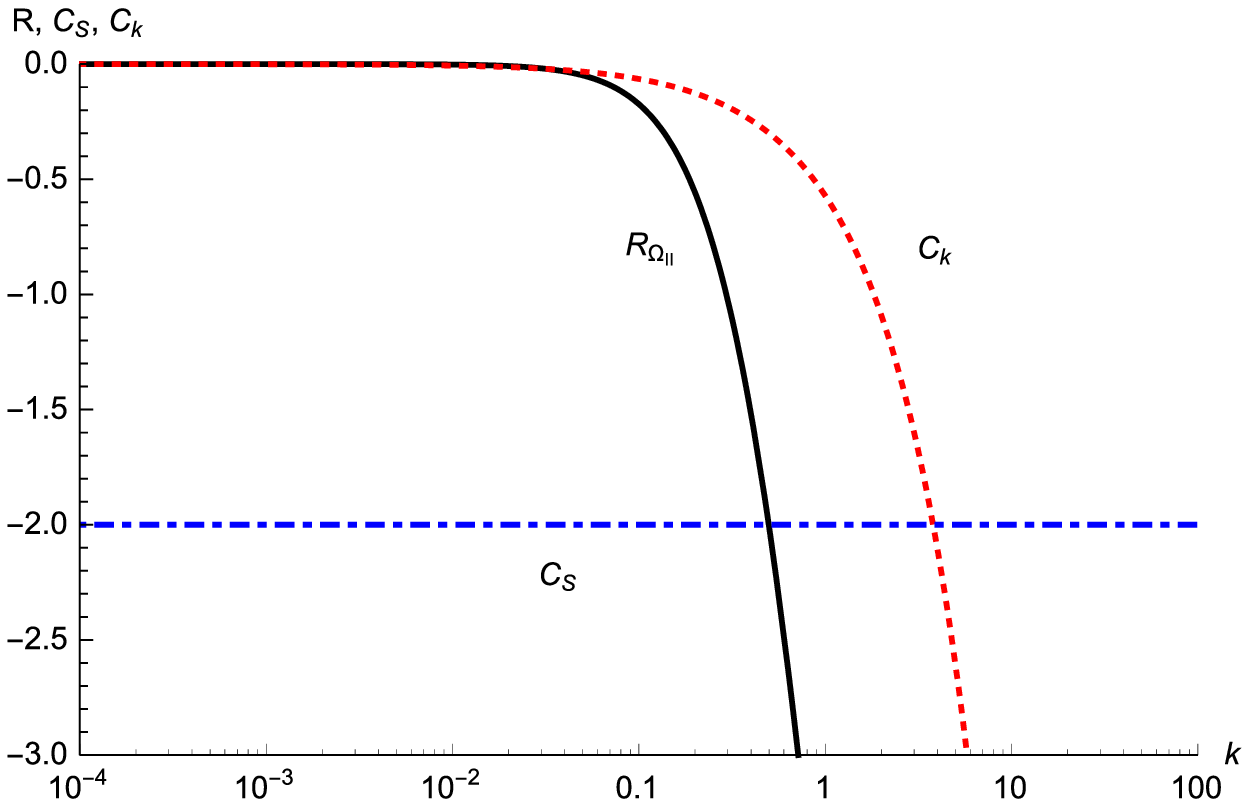}
\includegraphics[width=0.48\textwidth]{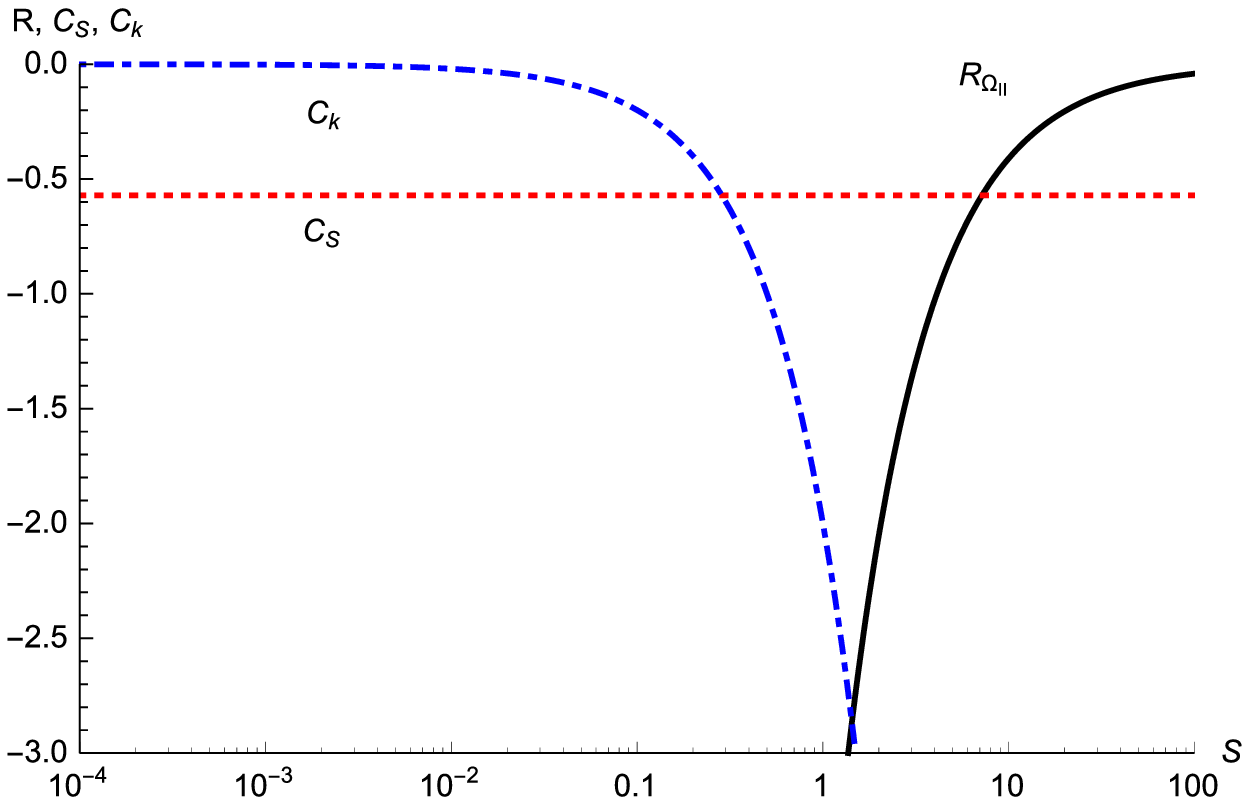}
\caption{\label{fig:Quevedo_runG}(Left) Ricci scalar $R_{\Omega_{II}}$ and heat capacities $C_{k}$ and $C_S$ for various $k$ at fixed $S=1$. (Right) Ricci scalar $R_{\Omega_{II}}$ and heat capacity $C_{k}$ and $C_S$ for various $S$ at fixed $k=1$.}
\end{figure}

\begin{figure}
\includegraphics[width=0.48\textwidth]{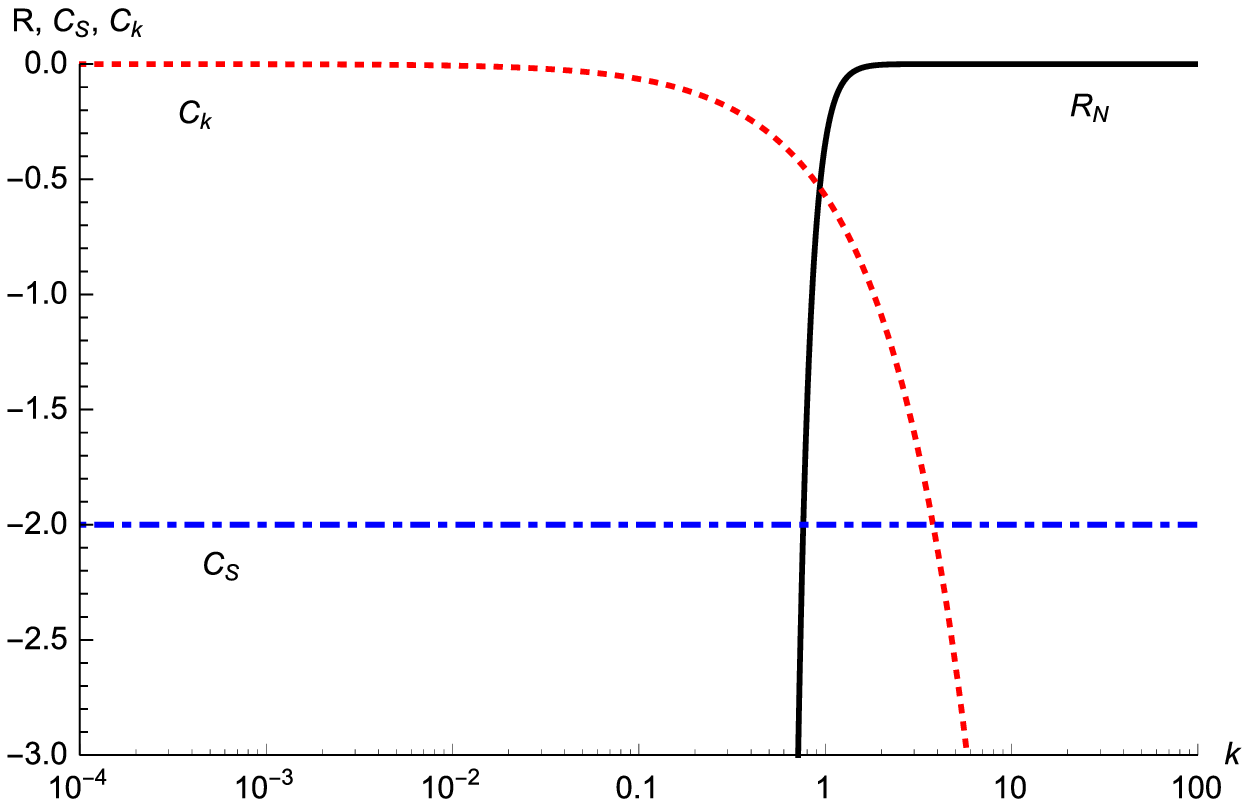}
\includegraphics[width=0.48\textwidth]{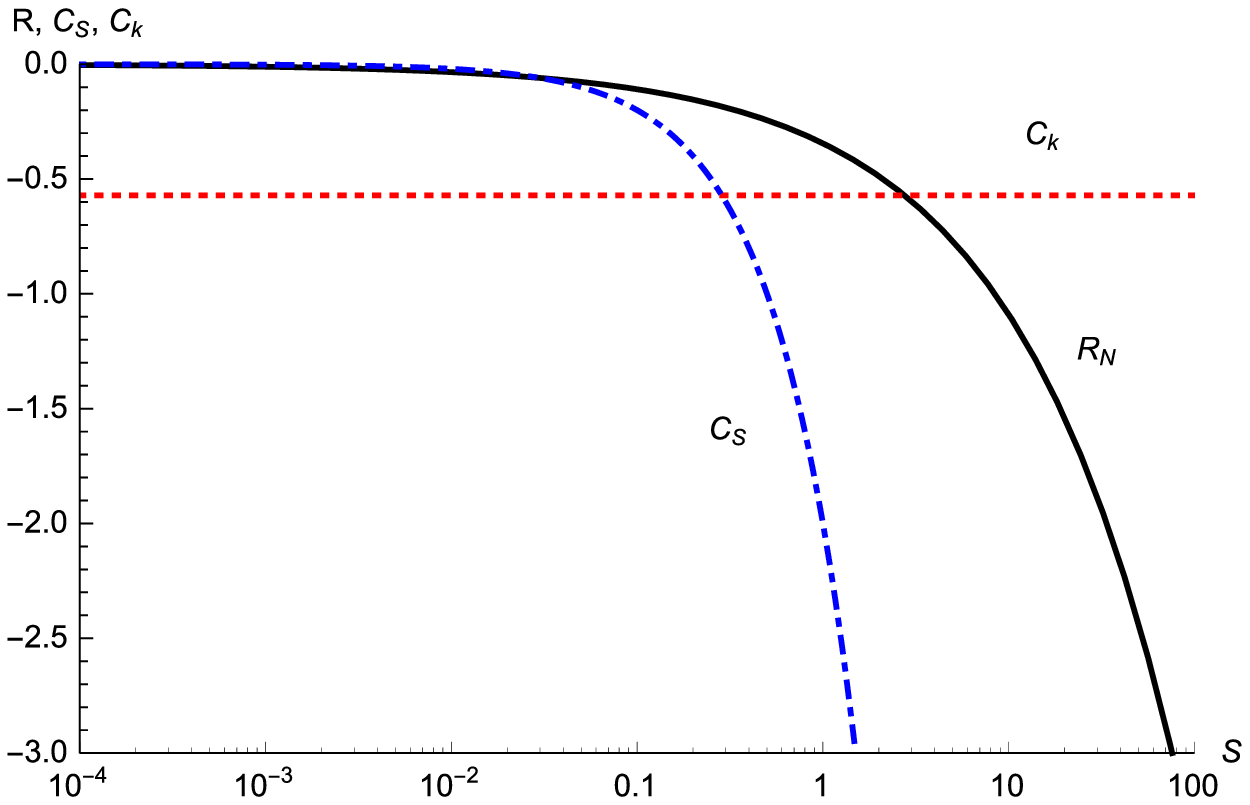}
\caption{\label{fig:HPEM_runG}(Left) Ricci scalar $R_N$ and heat capacities $C_{k}$ and $C_S$ for various $k$ at fixed $S=1$. (Right) Ricci scalar $R_N$ and heat capacity $C_{k}$ and $C_S$ for various $S$ at fixed $k=1$.
 }
\end{figure}

We first regard the Newton's gravitational constant $G$ as a free parameter and the Bekenstein-Hawking entropy as a function of both $G$ and mass $M$, that is
\begin{equation}\label{eqn:entropy_function}
S(G,M) = \frac{4\pi G M^2}{c^4}.
\end{equation}
The idea of time-varying $G$ was started long ago by Dirac to explain the coincidence between two large dimensionless numbers of ${\cal O}(10^{39})$ \cite{Dirac:1937}.  It appeared later in the Weinberg's asymptotic saftey for quantum gravity, which is based on the assumption that there exists a nontrivial fixed point for the renormalization group flow of $G$ \cite{Weinberg:1979}.  Now we compute the Ruppeiner metric, which is given by he negative Hessians of the entropy function (\ref{eqn:entropy_function}):
\begin{eqnarray}
ds^2_R =&& -S_{MM}dMdM - 2S_{MG}dMdG -S_{GG}dGdG\nonumber\\
=&& -\frac{8\pi G}{c^4} dM^2 -\frac{16\pi M}{c^4} dMdG,
\end{eqnarray}
where $S_{XY}$ denotes the differential $\partial_X\partial_Y S$.  It can be shown that Ruppeiner metric is flat.  Now we rewrite the ADM mass as a function of entropy and Newton constant, that is 
\begin{equation}
M(S,G)=\frac{c^2}{2}\sqrt{\frac{S}{\pi G}},
\end{equation}
which is identified as the internal energy of the black hole.  The Hessians of the internal energy gives rise to the Weinhold metric:
\begin{eqnarray}
ds^2_W =&& M_{SS}dSdS + 2M_{SG}dSdG + M_{GG}dGdG\nonumber\\
=&& -\frac{c^2}{8\sqrt{\pi}}\frac{1}{\sqrt{S^3G}}dS^2 -\frac{c^2}{4\sqrt{\pi}}\frac{1}{\sqrt{SG^3}} dSdG + \frac{3c^2}{8\sqrt{\pi}}\sqrt{\frac{S}{G^5}}dG^2, 
\end{eqnarray}
where $M_{XY}$ denotes the differential $\partial_X\partial_Y M$.  It can be shown that the Weinhold metric is also flat.  
Now we move to the Quevedo metric, which is given by\cite{Quevedo:2008xn} 
\begin{eqnarray}
ds^2_Q =&& \Omega (-M_{SS}dSdS + M_{GG}dGdG) \nonumber\\
=&& \Omega (\frac{c^2}{8\sqrt{\pi}}\frac{1}{\sqrt{S^3G}}dS^2 + \frac{3c^2}{8\sqrt{\pi}}\sqrt{\frac{S}{G^5}}dG^2)
\end{eqnarray}
with some conformal factor $\Omega$.  In the Quevedo metric of case I, one has the choice
\begin{equation}
\Omega_I = SM_S + GM_G = \frac{c^2}{4\sqrt{\pi}}\sqrt{\frac{S}{G}}+(-\frac{c^2}{4\sqrt{\pi}}\sqrt{\frac{S}{G}})=0
\end{equation}
or 
\begin{equation}
\Omega_{II} = SM_S = \frac{c^2}{4\sqrt{\pi}}\sqrt{\frac{S}{G}}
\end{equation}
for the case II.  We remark that there could exist infinite number of Legendre-invariant metrics and the above choices were picked up for simplicity.  Another new metric proposed by Hendi et al \cite{Hendi:2015rja}, denoting HPEM metric, claimed to remove the unwanted poles in the thermal Ricci scalar.  This corresponds to the choice of conformal factor $\Omega_N$:
\begin{equation}
\Omega_N = S\frac{M_S}{M_{GG}^3} = \frac{128\pi}{27c^4}\frac{G^7}{S}
\end{equation}
However, all Quevedo and HPEM metric are flat and have vanishing Ricci scalar.  This seems to imply that there is no thermodynamic interaction.  To have a nontrivial Ricci scalar, we assume a toy model for running Newton constant\cite{Falls:2012nd}:
\begin{equation}
\frac{1}{G(k)} = \frac{1}{k} + \frac{1}{G_{\infty}}, 
\end{equation} 
where $k$ is the energy scale.  The finite $G_{\infty}$ implies a vanishing Newton constant at IR and asymptotically UV safe theory of gravity.   Now we regard energy scale $k$, instead of $G$, as a free parameter and calculate various thermodynamic metric.   That is, we consider the entropy function $S=S(M,G(k))=S(M,k)$ for the Ruppeiner metric.  The Ricci scalar for the conformally flat Ruppeiner metric is found to be
\begin{equation}\label{eqn:Ricci_regular}
R_R = -\frac{c^4}{16\pi G_{\infty} M^2},
\end{equation}
which is independent of $k$.  The Ricci scalar for the Weinhold metric reads
\begin{equation}\label{eqn:Ricci_Weinhold_regular}
R_W = -\frac{\sqrt{\pi}}{c^2G_{\infty}}\sqrt{\frac{G(k)^3}{S}},
\end{equation}
which vanishes at IR and approaches constant at UV.  The heat capacity for constant $k$ and constant $S$ are computed respectively
\begin{equation}
C_k = -2S, \qquad C_S = -\frac{2k(k+G_\infty)}{4k+3G_\infty}.
\end{equation}
We remark that the heat capacity $C_k$ is explicitly independent on the scale $k$ and its negative value, which signals the instability, agrees with the conventional Schwarzschild black hole with constant $G$.  We plot them in the figure \ref{fig:Weinhold_runG}.  The Ricci scalars for both Ruppeiner and Weinhold metric remain negative for finite $S$ and $k$.  It becomes divergent as $M\to 0$ and vanished at IR limit ($M\to \infty$ or $k\to 0$).
While the Quevedo metric of case I turns out to be flat, the Quevedo metric of case II has a running Ricci scalar:
\begin{equation}
R_{\Omega_{II}}=-\frac{64\pi G_\infty k^2}{c^4 S (4k+3G_\infty )^2}
\end{equation}
which takes the following form at UV limit
\begin{equation}
\lim_{k\to \infty}R_{\Omega_{II}} = -\frac{4\pi G_\infty}{c^4 S}.
\end{equation}
We remark that both $R_R$ and $R_{\Omega_{II}}$ at UV limit are inversely proportional to $C_k$ if area law is true for all scales.  At last, the Ricci scalar for the HPEM metric reads
\begin{equation}
R_{N} = -\frac{c^2G_\infty S (72k^2+116kG_\infty+33 G_\infty^2)}{16\sqrt{\pi}k^6 (k+G_\infty)^4}\sqrt{\frac{G(k)}{S}},
\end{equation}
which vanishes at UV limit and scales as $-|C_k|^{1/2}$ for fixed $k$.  We plot them in the figure \ref{fig:Quevedo_runG} and \ref{fig:HPEM_runG}
\section{Regular Schwarzschild-like black hole}
Now we inspect various thermodynamic metric for a regular black hole which has the Schwarzschild metric from outset but a singularity-free interior.  It was first proposed by Hayward\cite{Hayward:2005gi} and recently served as a toy model to answer Hawking's resolution to the firewall and information paradox\cite{Frolov:2014jva}.  The metric reads:
\begin{equation}
g_{tt} = g^{rr} = 1-\frac{2GMr^2}{r^3+2\l^2GM},
\end{equation}
which behaves like the ordinary Schwarzschild black hole at large distance but a de Sitter space for $r \ll r_+$.  Now regarding $\l$ to be a new degree of freedom, we have the internal energy
\begin{equation}
M(S,\l)=\frac{\sqrt{S^3}}{2\sqrt{\pi}(S-\pi \l^2)}.
\end{equation}
The Hessian of $M$ gives the Weinhold metric:
\begin{equation}
ds^2_W = -\frac{S^2-6\pi\l^2 S-3\pi^2\l^4}{8\sqrt{\pi S}(S-\pi\l^2)^3} dM^2 - \frac{\sqrt{\pi S^3}\l (S+3\pi\l^2)}{S(S-\pi\l^2)^3} dM d\l + \frac{\sqrt{\pi S^3}(S+3\pi\l^2)}{(S-\pi\l^2)^3} d\l^2.
\end{equation}
which is conformally flat with Ricci scalar
\begin{equation}
R_W= \frac{2\sqrt{\pi}(S^2+6\pi\l^2S-15\pi^2\l^4)}{\sqrt{S}(S+3\pi\l^2)(S-3\pi\l^2)^2}
\end{equation} 
and the heat capacities evaluated at fixed $\l$ or fixed $S$ respectaively read
\begin{equation}
C_{\l} = -\frac{2S(S-\pi\l^2)(S-3\pi\l^2)}{S^2-6\pi\l^2 S-3\pi^2\l^4},\qquad C_S = -\l\frac{S-\pi\l^2}{S+3\pi\l^2}.
\end{equation}
We remark that a pole of $R_W$, $S=3\pi\l^2$, coincides with a zero of $C_{\l}$ at the extremal limit $r_+=\sqrt{3}\l$.    Furthermore, the ratio $\gamma \equiv C_S/C_{\l}$ has the same poles as that of Ricci scalar and removes the unphysical zero at $S=\pi \l^2$.  The heat capacity $C_{\l} \to -2S$ at the limit $S \gg \l^2$, as expected from the Schwarzschild black hole.

\begin{figure}
\includegraphics[width=0.48\textwidth]{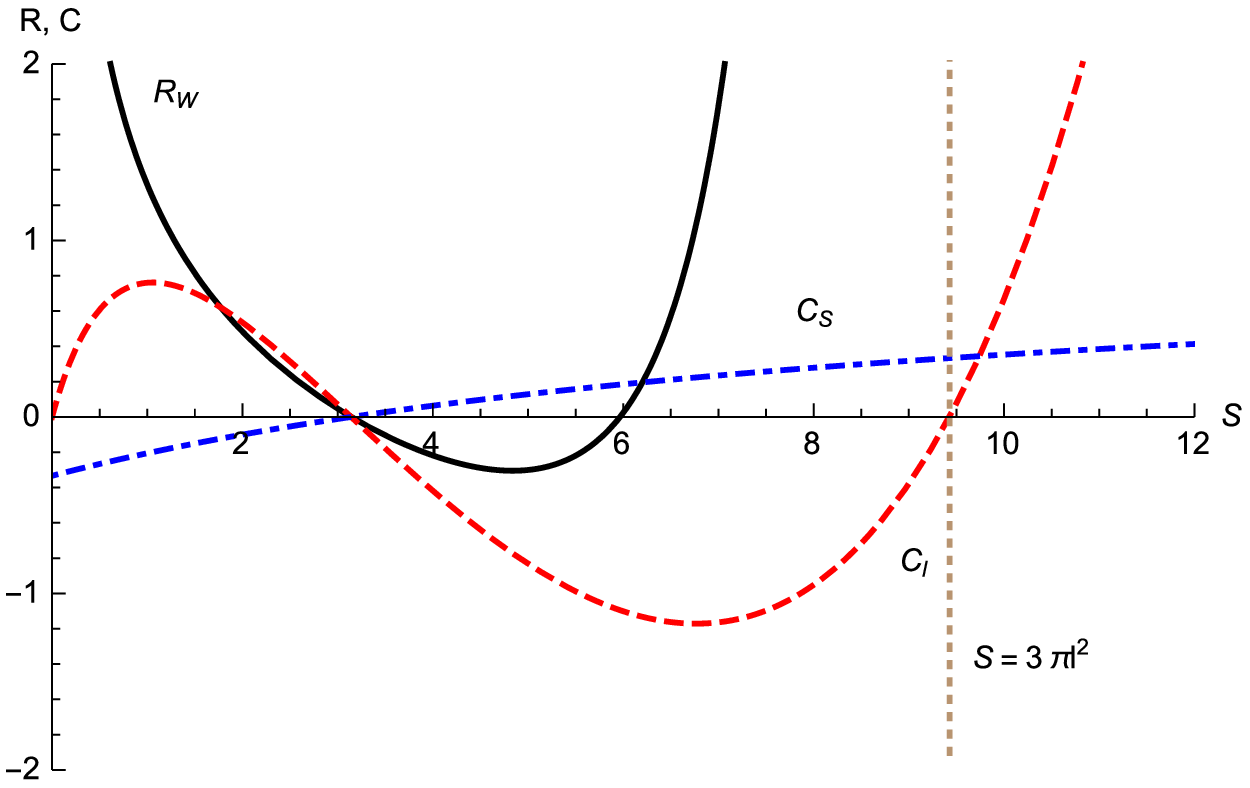}
\includegraphics[width=0.48\textwidth]{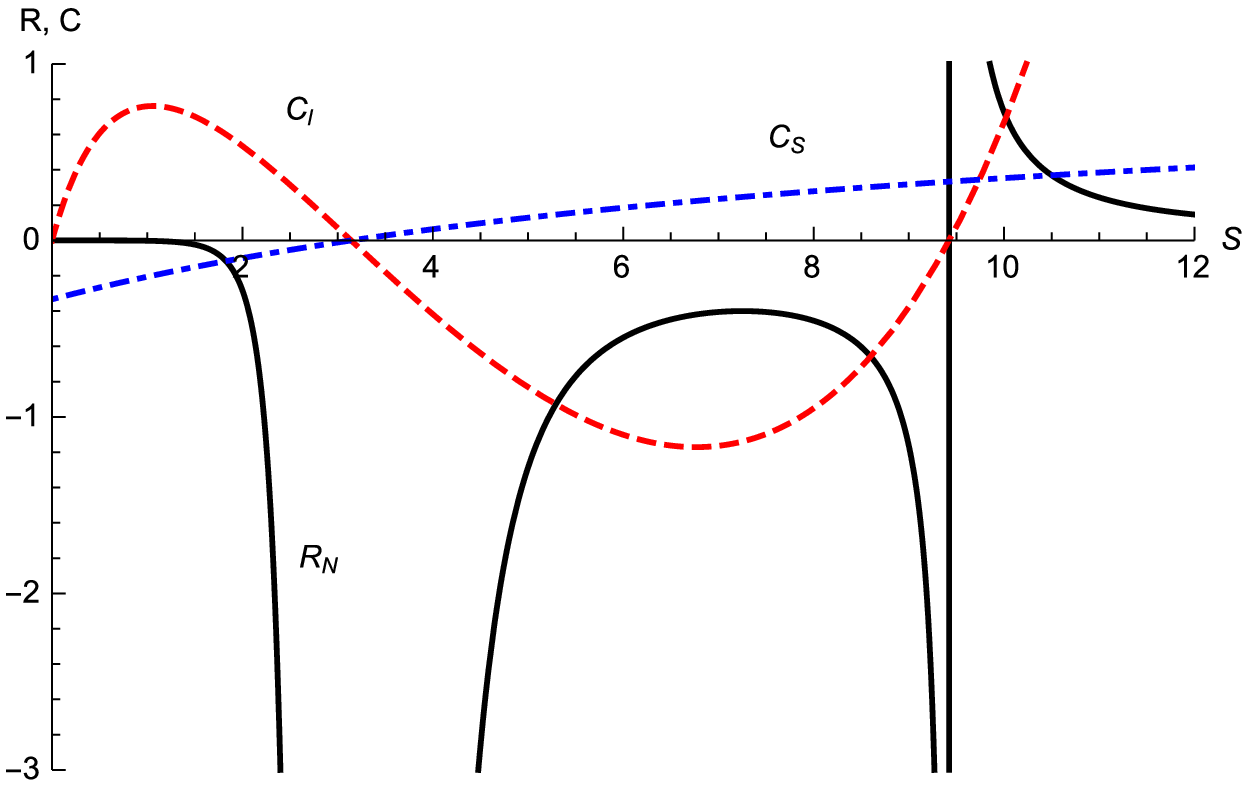}
\caption{\label{fig:Ricci_regular} (Left) Ricci scalar $R_W$ for the Weinhold metric and heat capacities $C_{\l}$ and $C_S$.  Here we set $\l=1$ for all simulation. (Right) Ricci scalar $R_N$ for the HPEM metric and heat capacity $C_{\l}$ and $C_S$.}
\end{figure}

\begin{figure}
\includegraphics[width=0.48\textwidth]{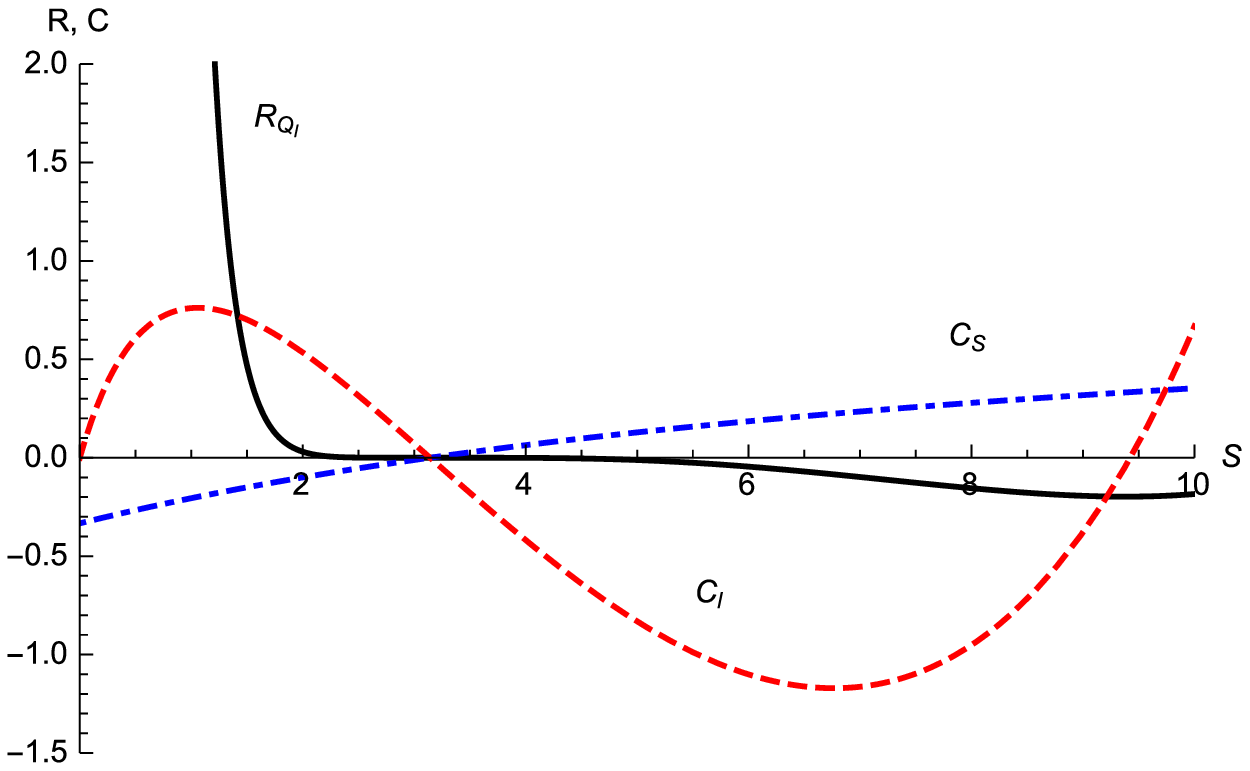}
\includegraphics[width=0.48\textwidth]{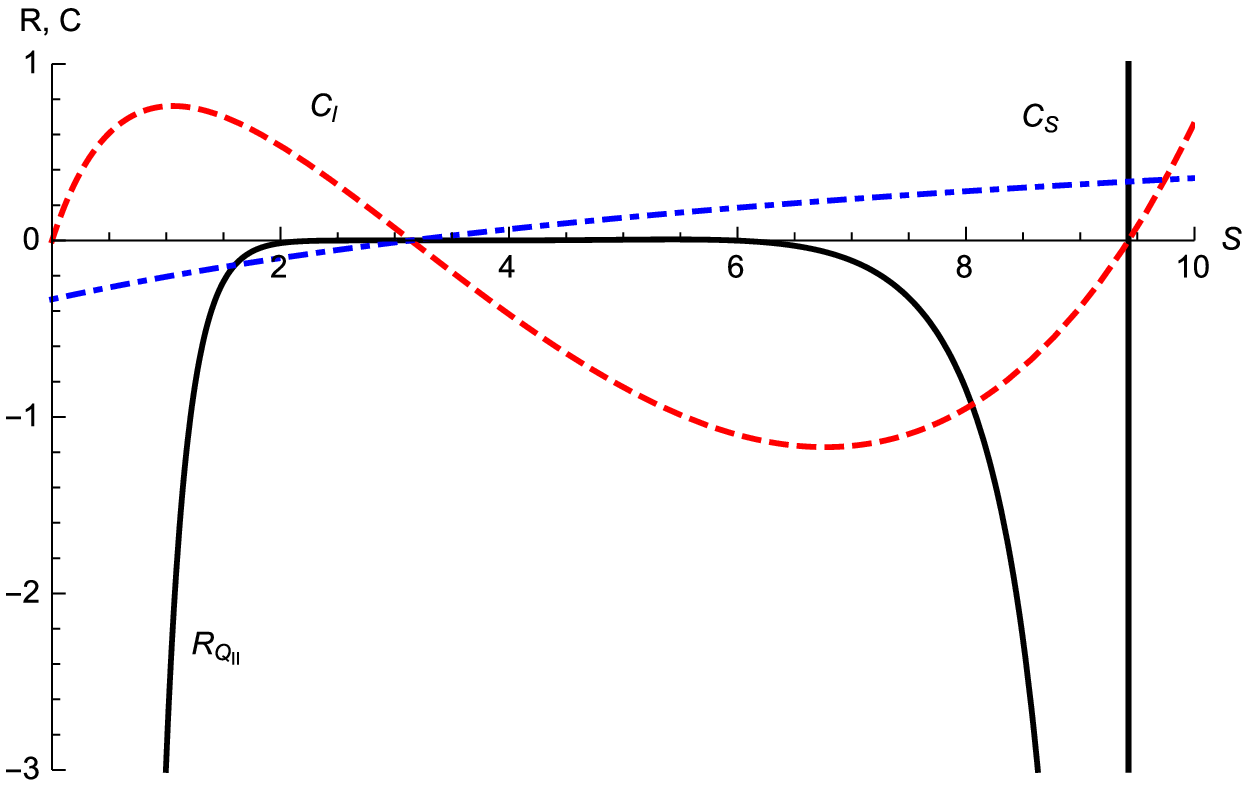}
\caption{\label{fig:Ricci_Q_regular} (Left) Ricci scalar $R_{\Omega_I}$ for the Quevedo model I and heat capacities $C_{\l}$ and $C_S$. (Right) Ricci scalar $R_{\Omega_{II}}$ for the Quevedo model II and heat capacities $C_{\l}$ and $C_S$. }
\end{figure}

Now we study the Quevedo metric.  The conformal factors for the case I and II are
\begin{equation}
\Omega_I = \frac{\sqrt{S^3}(S+\pi\l^2)}{4\sqrt{\pi}(S-\pi\l^2)^2},\qquad \Omega_{II} = \frac{\sqrt{S^3}(S-3\pi\l^2)}{4\sqrt{\pi}(S-\pi\l^2)^2}
\end{equation}
respectively.  The corresponding Ricci scalar for case I is lengthy to be skipped here, but its denominator reads
\begin{equation}
\text{demoninator}( R_{Q_I})=S^2(S+\pi\l^2)^3(S+3\pi\l^2)^2(S^2-6\pi\l^2 S-3\pi^2\l^4)^2.
\end{equation}
Although there is no pole which agrees with the extremal limit, it poccesses the same positive pole as that of $C_{\l}$, i.e. $S=(3+2\sqrt{3})\pi\l^2$.  The denominator of Ricci scalar for case II reads
\begin{equation}
\text{demoninator}( R_{Q_{II}} )= S^2(S-3\pi\l^2)(S+3\pi\l^2)(S^2-6\pi\l^2-3\pi^2\l^4)^2
\end{equation}
and the desired pole, $S=3\pi\l^2$, is one of them.  As to the HPEM metric, the denominator for its Ricci scalar reads
\begin{equation}
\text{demoninator}( R_N )= (S-3\pi\l^2)(S-\pi\l^2)^4(S^2-6\pi\l^2-3\pi^2\l^4)^2,
\end{equation}
while it keeps the desired pole but replaces the pole $S=0$ by $S=\pi\l^2$ instead.
We remark that at the Schwarzschild limit $\l \to 0$, the Ricci scalar for different metrics read 
\begin{equation}
R_W = 2\sqrt{\frac{\pi}{S}},\quad R_{Q_I} = 0,\quad R_{Q_{II}} = \frac{32\pi}{S}, \quad R_N =200\sqrt{\frac{\pi^5}{S^5}}.
\end{equation}
We remark that beside the case I of Quevedo metric, Ricci scales are inversely proportional to $S$ to some power as $\l \to 0$.  We plot those Ricci scalars with heat capacities in the figure \ref{fig:Ricci_regular} and \ref{fig:Ricci_Q_regular}.  The nonvanishing Ricci scalar implies that strength of thermodynamic interaction increases as the black hole evaporates.  The divergance at the final stage suggests a phase transition where new degrees of freedom in the quantum gravity would play an important role.

\section{Noncommutative geometry inspired Schwarzschild black hole}
In the noncommutative geometry, a smear source was proposed to resolve the singularity at the center of Schwarzschild black hole\cite{Nicolini:2005vd}:
\begin{equation}
M(\sqrt{\frac{S}{\pi}},\theta) = \frac{\sqrt{S}}{4\gamma(\frac{3}{2},\frac{S}{4\pi\theta})},
\end{equation}
where the imcomplete gamma function $\gamma(a,x)=\int_0^x{t^{a-1}e^{-t}dt}$.
\begin{figure}
\includegraphics[width=0.48\textwidth]{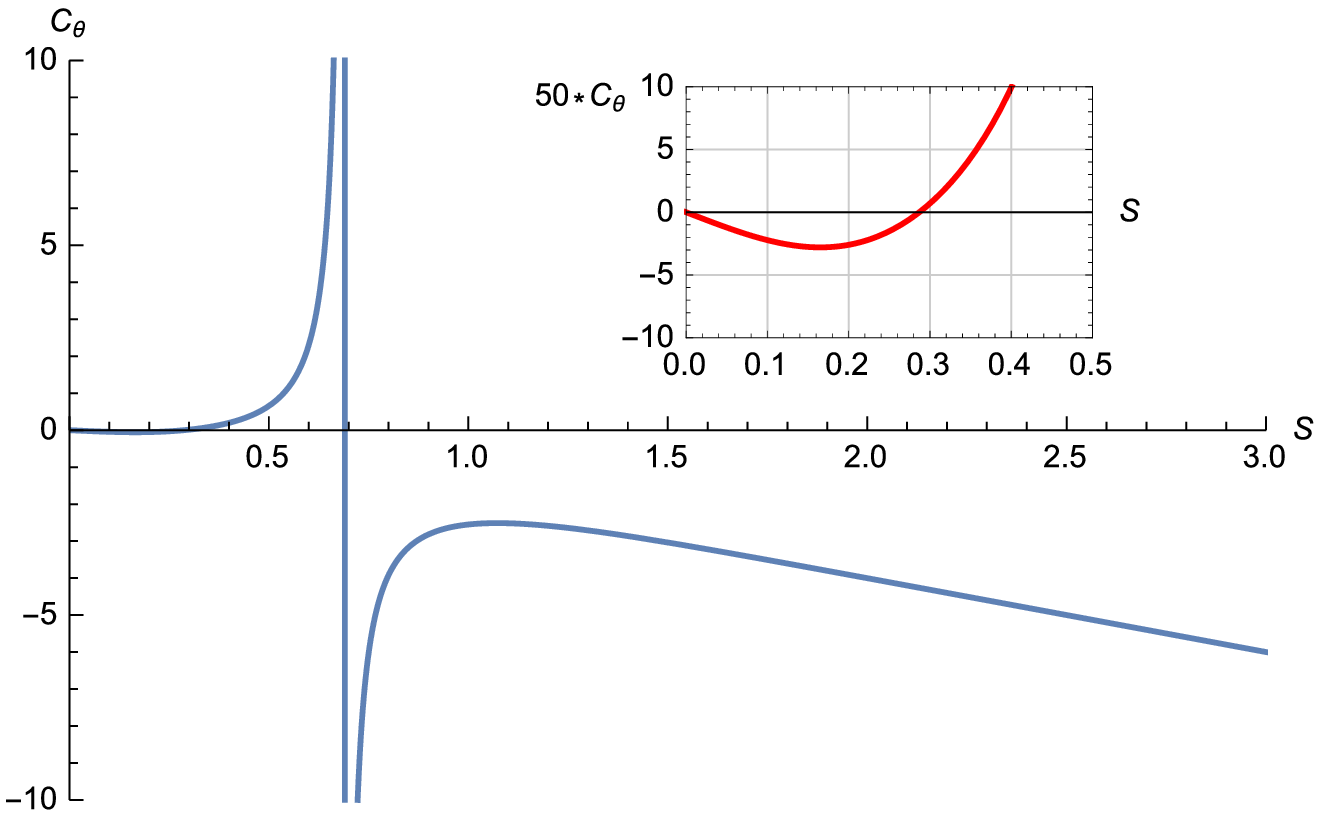}
\includegraphics[width=0.48\textwidth]{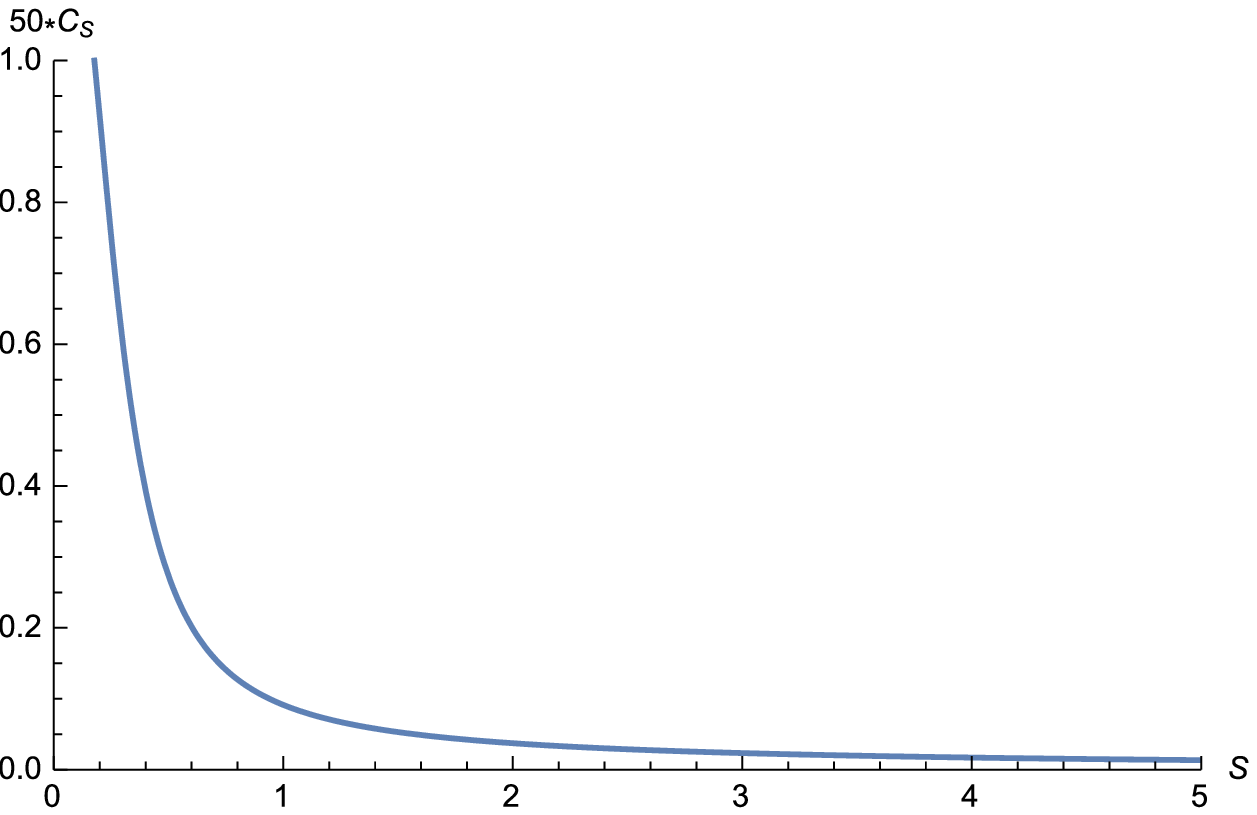}
\caption{\label{fig:Capacity_NC} (Left) Heat capacity $C_\theta$ for fixed $\theta=0.01$  The inset plot shows a zero at small $S$. (Right) Heat capacity $C_S$ for different $S$. }
\end{figure}

\begin{figure}
\includegraphics[width=0.48\textwidth]{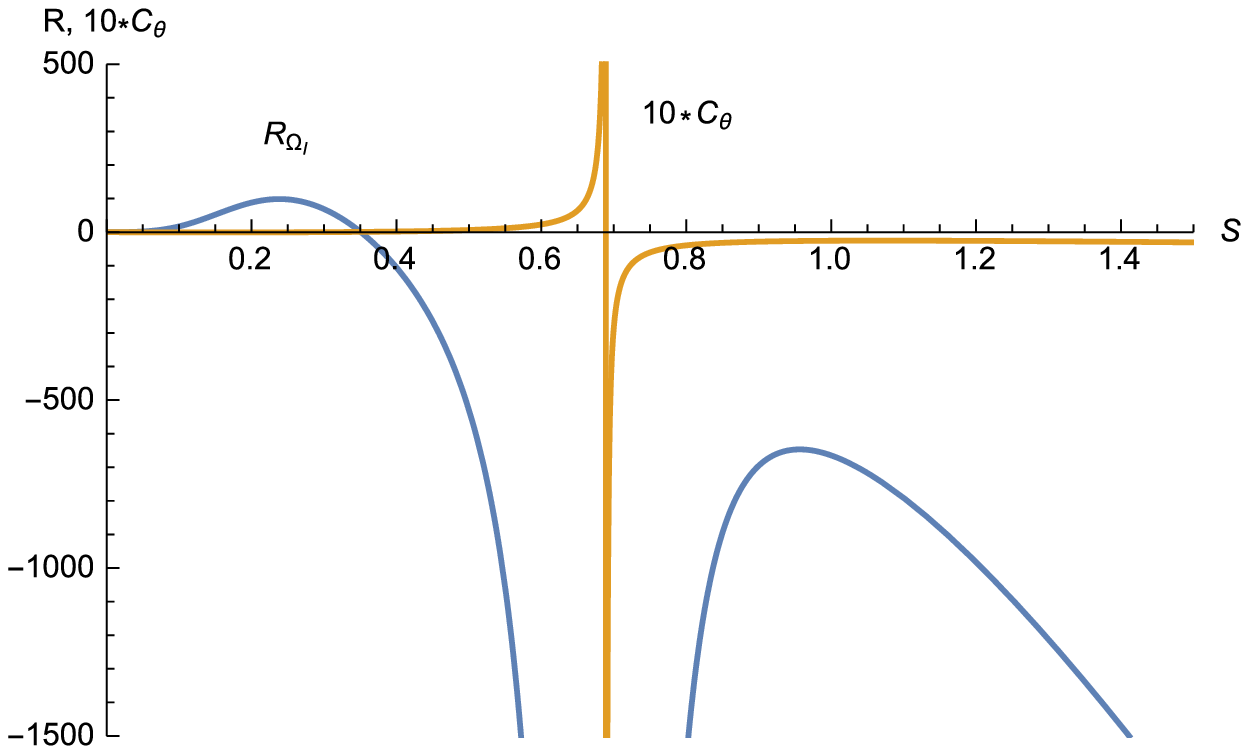}
\includegraphics[width=0.48\textwidth]{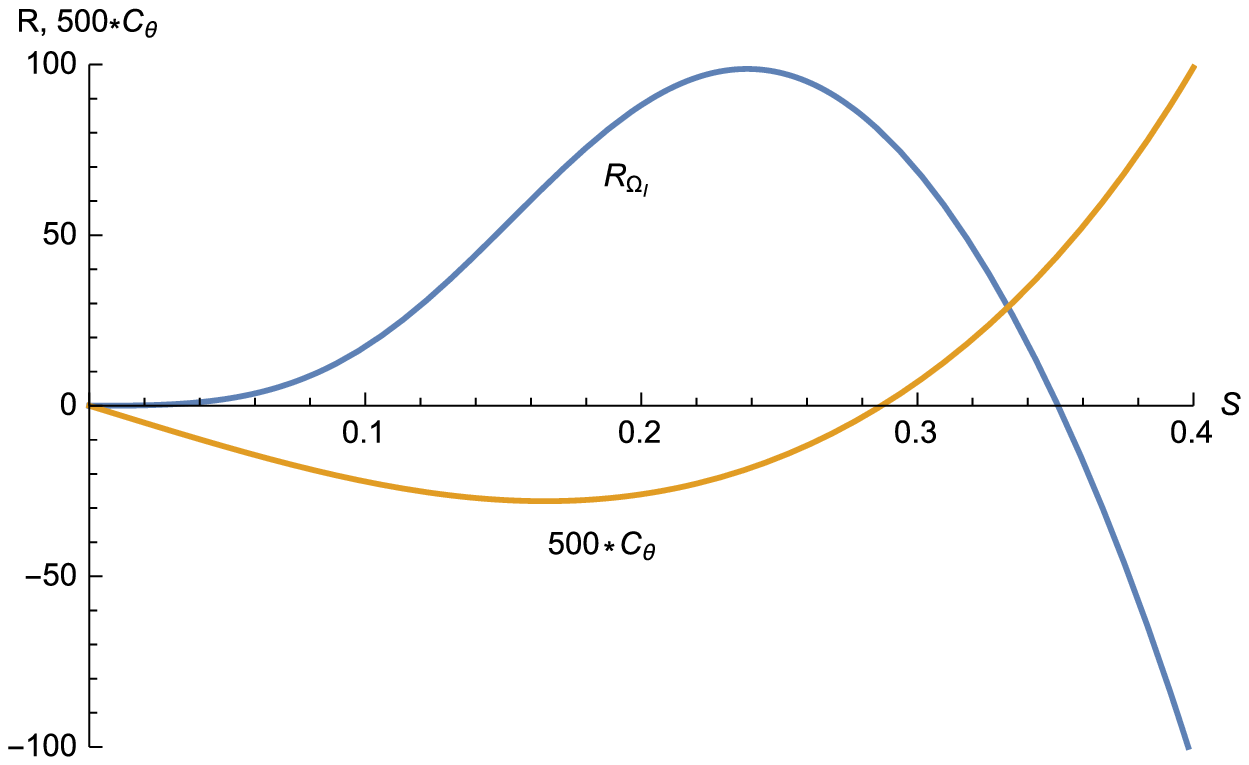}
\caption{\label{fig:Ricci_NC_Q1} (Left) Ricci scalar for Quevedo metric model I of NC black hole and heat capacity for constant $\theta$. (Right) Ricci scalar for Quevedo metric model I of NC black hole and heat capacity for constant $\theta$ at small $S$. }
\end{figure}

\begin{figure}
\includegraphics[width=0.48\textwidth]{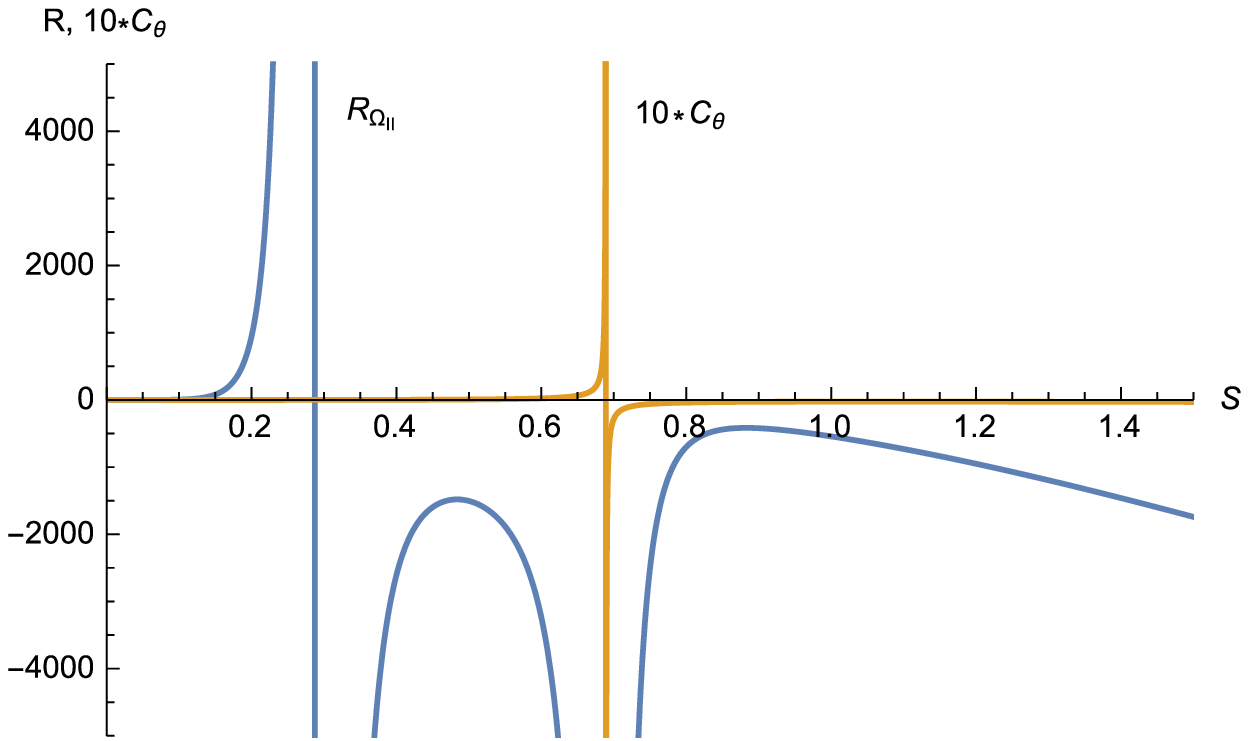}
\includegraphics[width=0.48\textwidth]{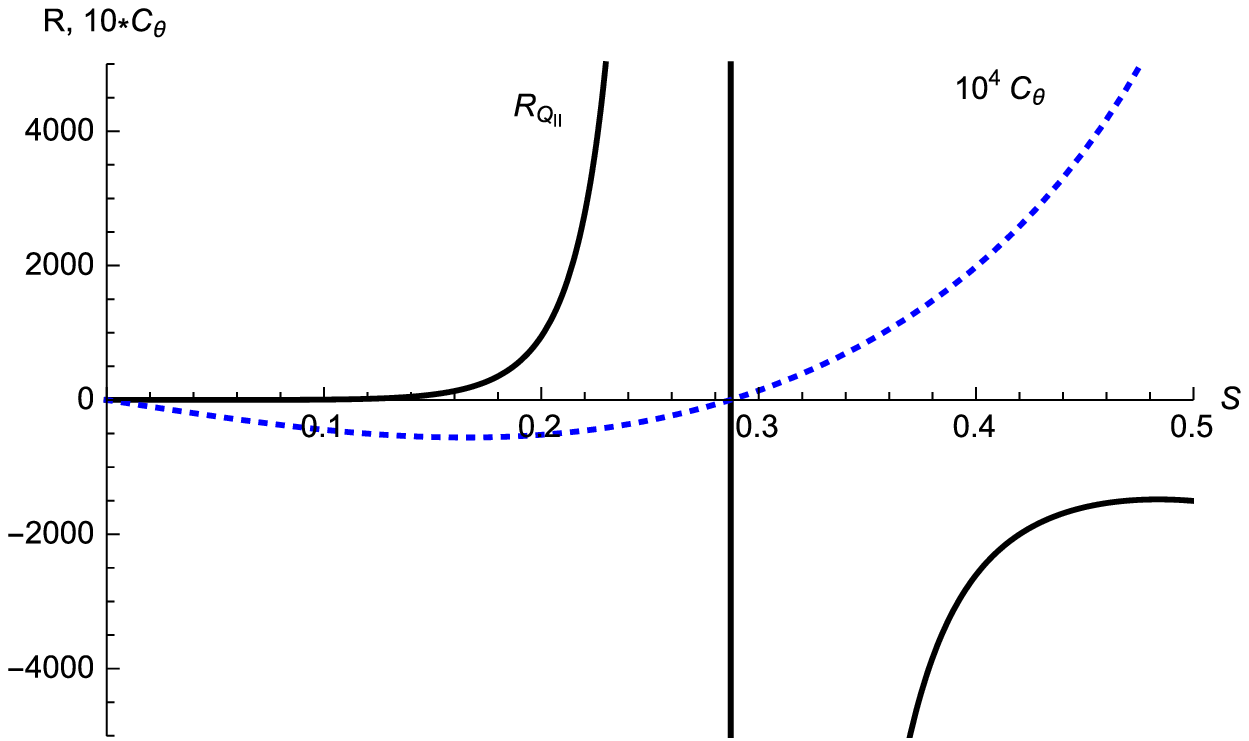}
\caption{\label{fig:Ricci_NC_Q2} (Left) Ricci scalar for Quevedo metric model II of NC black hole and heat capacity for constant $\theta$. (Right) Ricci scalar for Quevedo metric model II of NC black hole and heat capacity for constant $\theta$ at small $S$. }
\end{figure}

\begin{figure}
\includegraphics[width=0.48\textwidth]{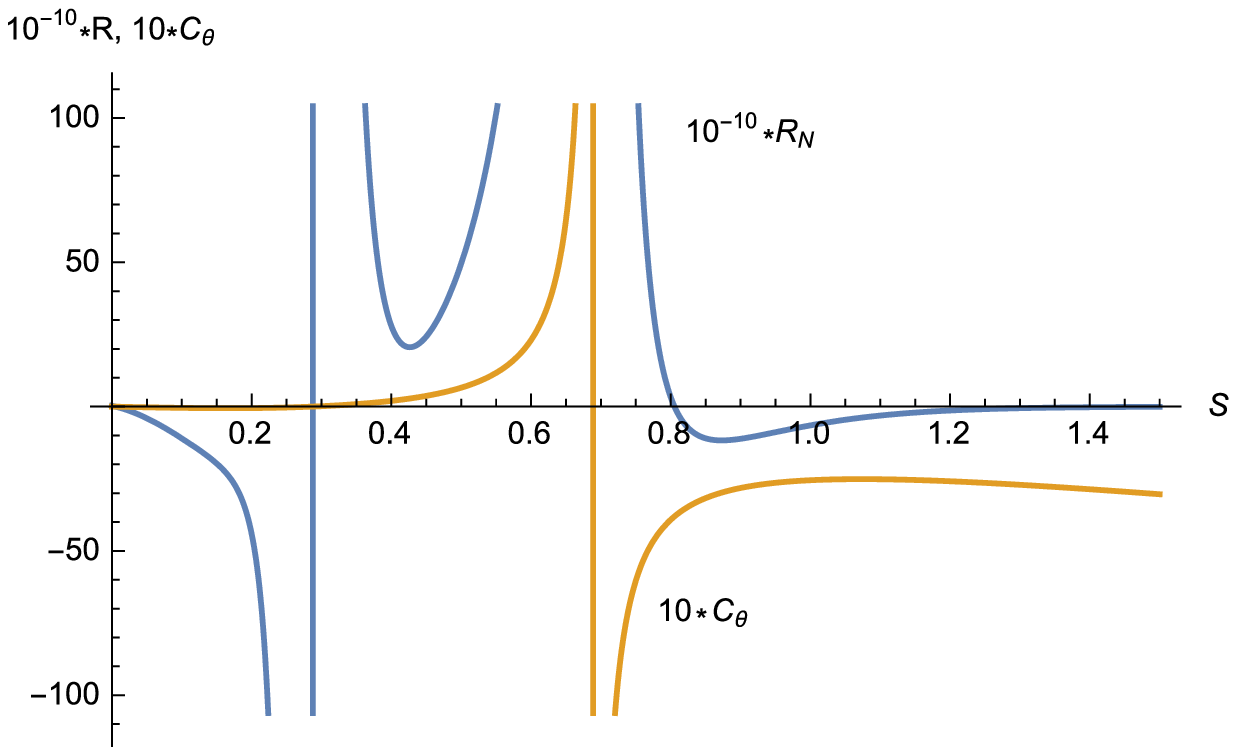}
\includegraphics[width=0.48\textwidth]{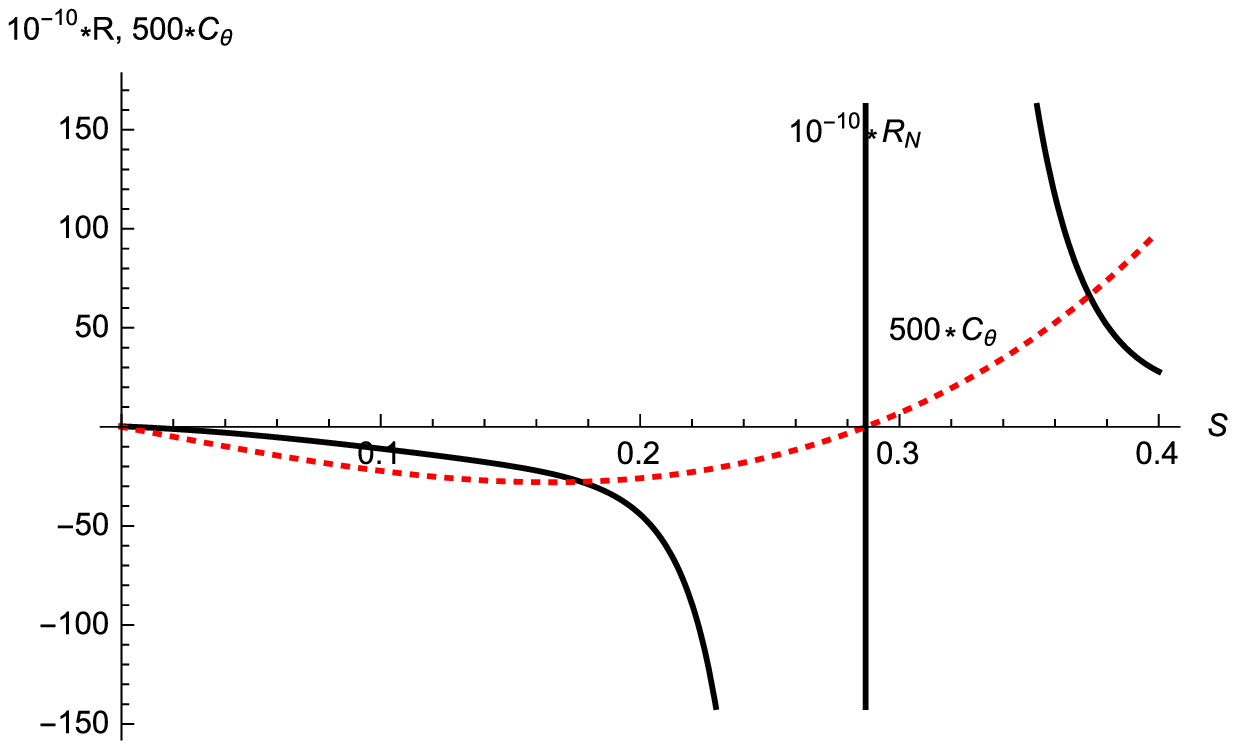}
\caption{\label{fig:Ricci_NC_HPEM} (Left) Ricci scalar for HPEM metric of NC black hole and heat capacity for constant $\theta$. (Right) Ricci scalar for HPEM metric of NC black hole and heat capacity for constant $\theta$ at small $S$. }
\end{figure}

Now regarding the noncommutative parameter $\theta$ as a free varaible, one can compute the lengthy Weinhold metric, wchih is flat.   Heat capacities $C_\theta$ at fixed $\theta$ and $C_S$ at fixed $S$ are plotted in the figure \ref{fig:Capacity_NC}.  We remark that $C_\theta$ has a zero at small $S$ and follows by a spike.  It approaches the line $C_\theta = -2S$ for large $S$, as expected for the Schwarzschild black hole.  On the other hand, $C_S$ monotoneously decreases with $S$.
Both models in the Quevedo metric are all conformally flat.  We plot their Ricci scalars together with heat capicities in the figure \ref{fig:Ricci_NC_Q1} and figure \ref{fig:Ricci_NC_Q2}.  We remark that the pole of $R_{\Omega_I}$ matches with the spike of $C_\theta$, but their zeros at small $S$ do not agree.  On the other hand, the spike of $R_{\Omega_{II}}$ agrees with the zero of $C_\theta$ at small $S$.  The Riccis scalar of HPEM metric $R_N$ is potted in the figure \ref{fig:Ricci_NC_HPEM} and the zero and spike of $C_\theta$ agrees with the spike and pole of $R_N$.

It is not straightforward to take $\theta \to 0$ limit for the thermal Ricci scalar.  However, if the above onshell limit is taken with the extremal condition $S=9\pi \theta$, it is found that 
\begin{equation}
R \sim -\frac{C}{\theta},
\end{equation}
where $C\simeq 0.72$ for the model I of Quevedo metric and $C\simeq 1.28\times 10^5$ for the model II.   Since the extremal limit can be regarded as a kind of vacuum (ground state) of noncommutative space for its zero temperature, it is entertaining to suggest the divergence of Ricci scalar at $\theta \to 0$ limit implies a phase transition from the noncommutative space to the commutative space.

\section{Shcwarzschild black hole with Tsallis-R{\`e}nyi entropy}

Bir{\'o} and Czinner \cite{Biro:2013cra} regarded the Hawking-Bekenstein entropy as the non-extensive Tsallis or R{\`e}nyi entropy.  The Tsallis entropy has been devised to fit the power-law tailed spectra, and the Schwarzschild black hole becomes thermally stable at a fixed temperature in a similar way as that in the anti-de Sitter space\cite{Biro:2013cra,Czinner:2015eyk}.  Here the R{\`e}nyi entropy reads, 
\begin{equation}\label{eqn:Renyi_entropy}
S_a(M) = \frac{1}{a}\ln(1+4\pi a M^2),
\end{equation}
where the Bekenstein-Hawking entropy is obtained at the Schwarzschild limit $a\to 0$.

Now we regard the entropy as a function of both mass and $a$, say $S(M,a)=S_a(M)$.  The Ruppeiner metric is conformally flat.  The Ricci scalar expands around the Schwarzschild limit as:
\begin{equation}
R_R = \frac{45}{256\pi M^2} + \frac{171}{1280}a - \frac{549 \pi}{5120} M^2 a^2 +\cdots.
\end{equation}
The Weinhold metric can be obtained from inverting (\ref{eqn:Renyi_entropy}) and the Ricci scalar expands around the Schwarzschild limit as:
\begin{equation}
R_W = -\frac{19}{16}\sqrt{\frac{\pi}{S}} + \frac{123}{160}\sqrt{\pi S} a - \frac{3721}{15360}\sqrt{\pi S^3}a^2 + \cdots.
\end{equation}

\begin{figure}
\includegraphics[width=0.48\textwidth]{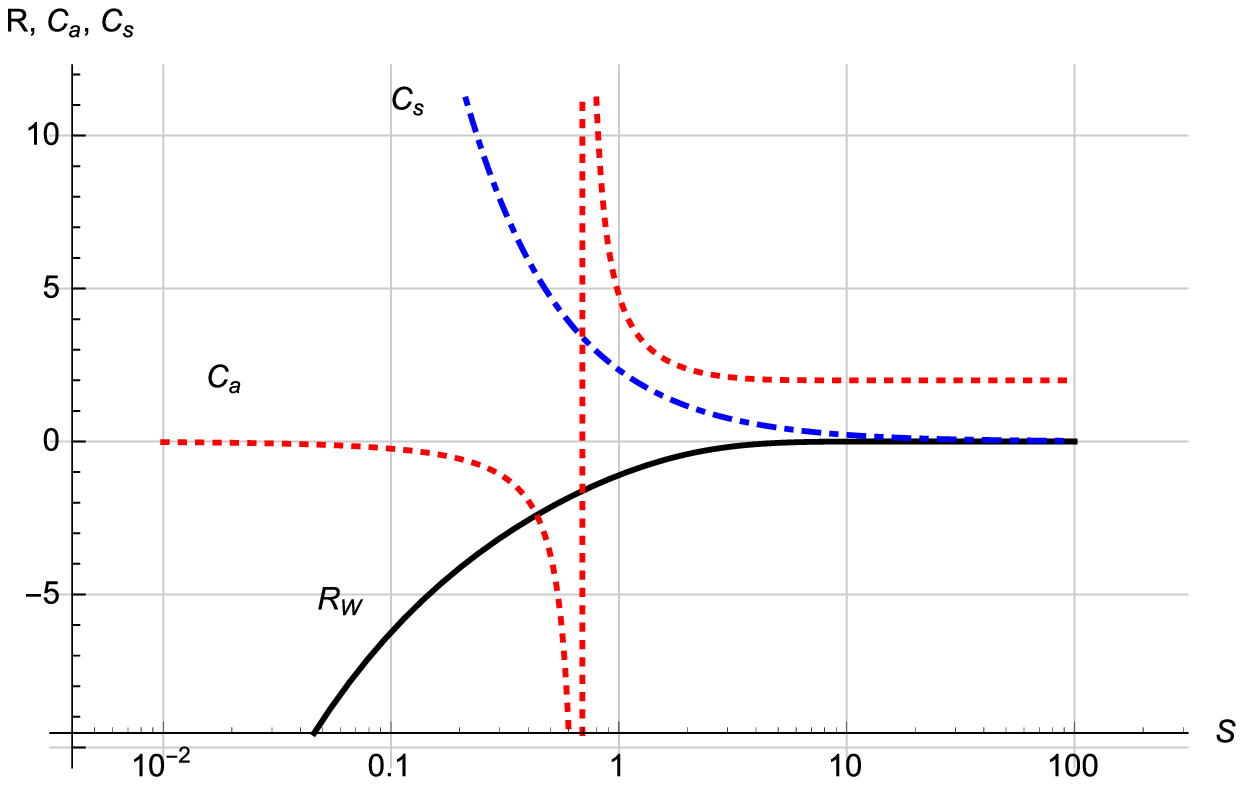}
\includegraphics[width=0.48\textwidth]{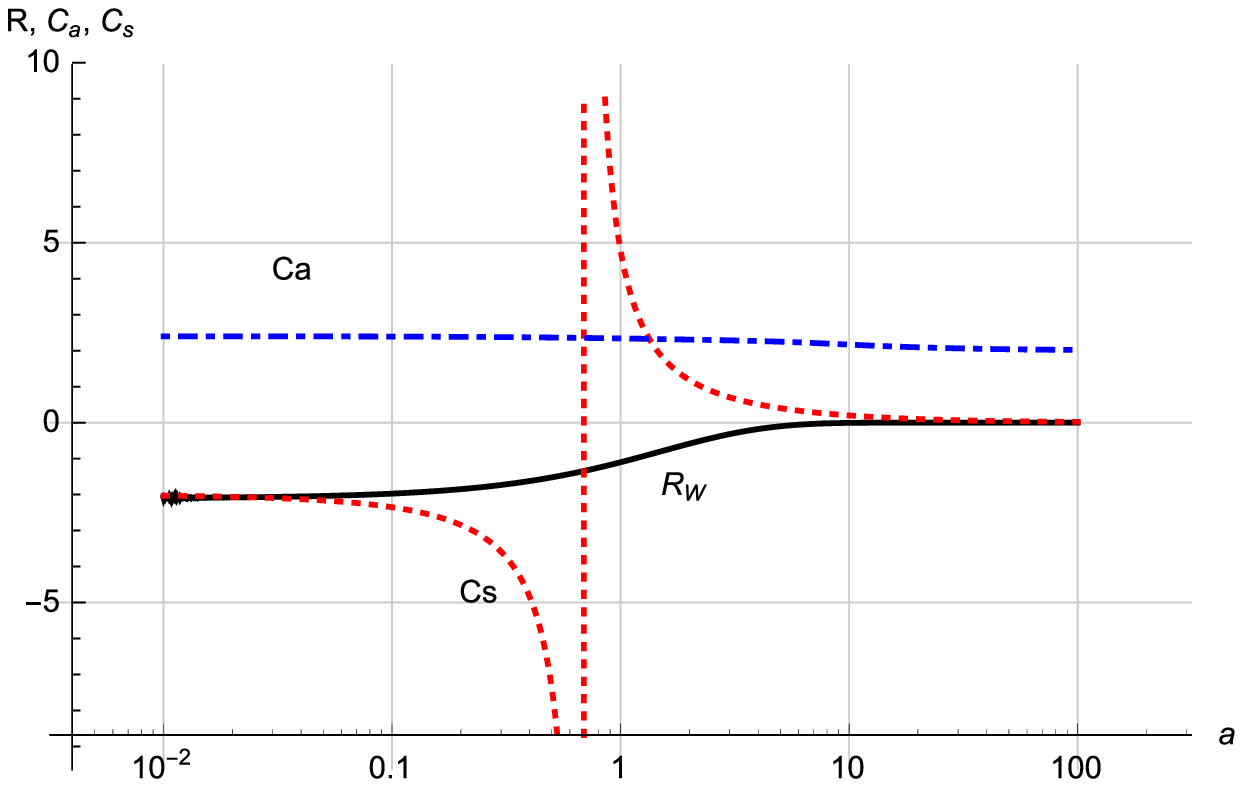}
\caption{\label{fig:Renyi_Weinhold} (Left) Ricci scalar for Weinhold metric for black hole with R{\`e}nyi entropy and heat capacity for constant $a$ and constant $S$ at fixed $a=1$. (Right) Ricci scalar for Weinhold metric for black hole with R{\`e}nyi entropy and heat capacity for constant $a$ and constant $S$ at fixed $S=1$. }
\end{figure}

We remark that the leading terms in $R_R$ and $R_W$ behave like those in (\ref{eqn:Ricci_regular}) and (\ref{eqn:Ricci_Weinhold_regular}) but with different coefficients.   We plot the Weinhold Ricci scalar and heat capacities against entropy and parameter $a$ respectively in the figure \ref{fig:Renyi_Weinhold}.

\begin{figure}
\includegraphics[width=0.48\textwidth]{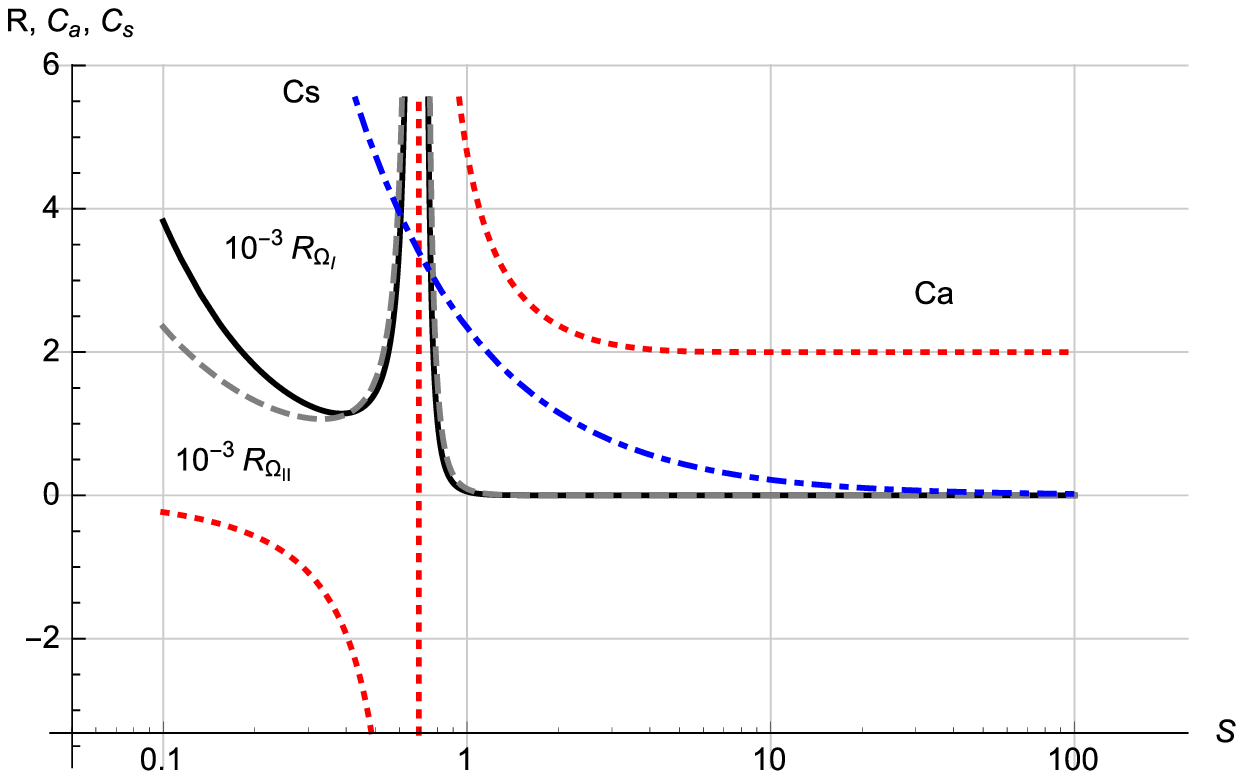}
\includegraphics[width=0.48\textwidth]{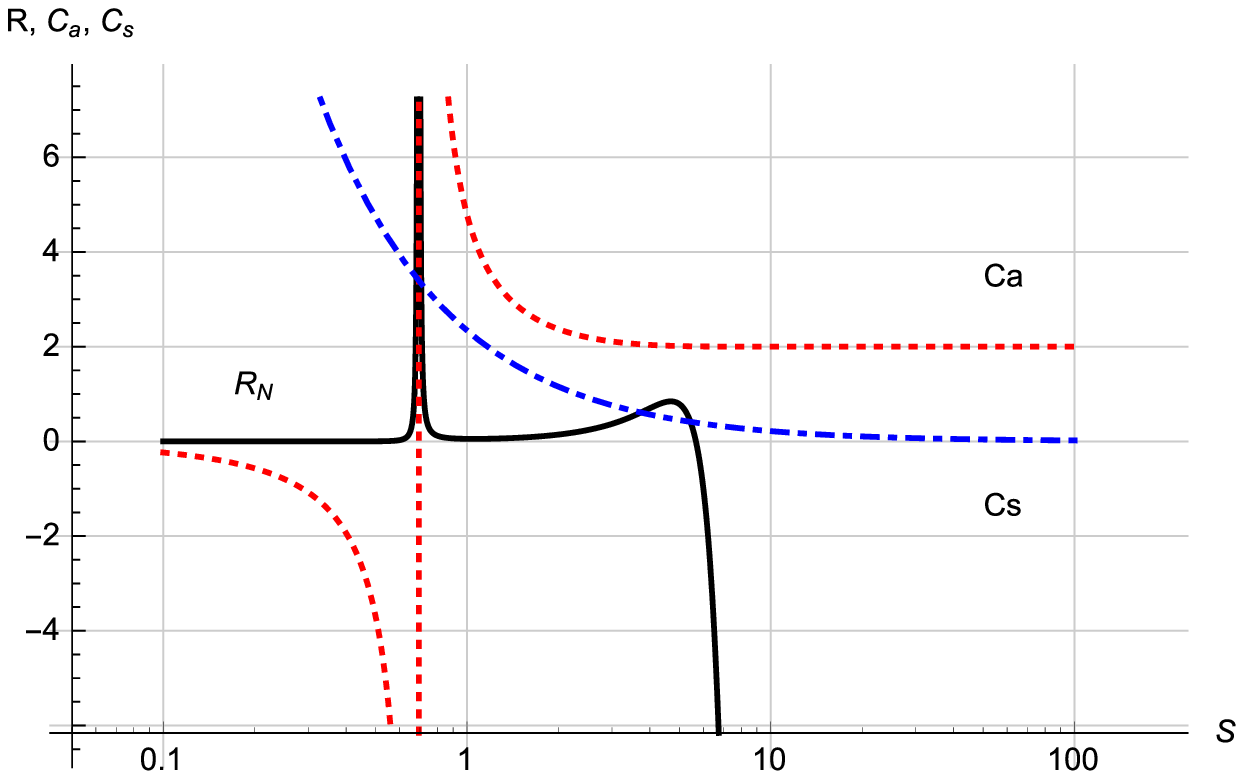}
\caption{\label{fig:Renyi_Quevedo} (Left) Ricci scalar for Quevedo metric for black hole with R{\`e}nyi entropy and heat capacity for constant $a$ and constant $S$ at fixed $a=1$. (Right) Ricci scalar for HPEM metric for black hole with R{\`e}nyi entropy and heat capacity for constant $a$ and constant $S$ at fixed $a=1$. }
\end{figure}

In the figure \ref{fig:Renyi_Quevedo}, we plot the Ricci scalars for Quevedo metric and HPEM metric.  The divergence of both Ricci scalars agree with the heat capacity $C_a$.  At the Schwarzschild limit, their Ricci scalars behave as follows:
\begin{equation}
R_{\Omega_I}= \frac{3456\pi}{25 S},\qquad R_{\Omega_{II}} = \frac{352\pi}{5S}, \qquad R_{N} = \frac{965}{27648}\sqrt{\frac{S^{13}}{\pi}}.
\end{equation}

\section{Discussion}
In this letter, we study the offshell thermal metric of Schwarzschild black hole by introducing a new degree of freedom, which could be the running Newton constant,  a cutoff scale for regular black hole, a noncommutative deformation, or the deformed parameter in the nonextensive Tsallis-R{\`e}nyi entropy.  The onshell thermal Ricci scalar of original Schwarzschild black hole is obtained by guage fixing this freedom.  In the table \ref{table:summary}, we summarize our result for different metric and various deformation.  In many cases, the Ricci scalar blows up at the final stage of evaporation where $S\to 0$.  This calls for a theory of quantum gravity to resolve the divergence.   For those effective theories with deformation, the Ricci scalar in the Quevedo or HPEM metric usually has a pole at finite entropy, which agrees with either a zero or pole in the heat capacity.  This implies the existence of a phase transition at some UV scale due to emerged degrees of freedom in quantum effects.

\begin{table}
\begin{center}
    \begin{tabular}{ || c | c | c | c | c | c  ||}
\hline 
    Theories & Ruppeiner & Weinhold & Quevedo I & Quevedo II & HPEM \\ \hline
    Running G\footnote{We also take the UV limit $k\to \infty$.} & $-S^{-1}$ & $-S^{-1/2}$ & $0$ & $-S^{-1}$ & $0$\\ \hline
    Regular BH &  -\footnote{We do not carry the computation in this letter.}  & $S^{-1/2}$ & $0$ & $S^{-1}$  & $S^{-5/2}$ \\ \hline
    Noncommutative BH\footnote{Extremal limit is taken before sending $\theta \to 0$.} & - & $-S^{-1}$ & $-S^{-1}$ & $-S^{-1}$ & $-S^{-1}$  \\ \hline
    R{\`e}nyi entropy &  $S^{-1}$ & $-S^{-1/2}$ & $S^{-1}$ & $S^{-1}$ & $S^{13/2}$ \\
\hline
    \end{tabular}
\end{center}
\caption{\label{table:summary}{\sl Onshell} thermal Ricci scalar at the Schwarzschild limit.  Here we assume the Bekenstein-Hawking area law is correct.}
\end{table}

\begin{acknowledgments}
The author is grateful to the package Riemannian Geometry \& Tensor Calculus and the software Mathematica.  This work is supported in parts by the Taiwan's Ministry of Science and Technology (grant No. 102-2112-M-033-003-MY4) and the National Center for Theoretical Science. 
\end{acknowledgments}


\end{document}